\documentclass[11pt,a4paper]{article}

\usepackage[T1]{fontenc}

\usepackage[utf8]{inputenc} \usepackage[T1]{fontenc}
\usepackage{tgtermes}
\usepackage{amssymb}
\usepackage{amsfonts}

\usepackage{tgtermes}
\usepackage{amsmath}
\usepackage{amsthm}  \usepackage{scalefnt,letltxmacro}
\LetLtxMacro{\oldtextsc}{\textsc}
\renewcommand{\textsc}[1]{\oldtextsc{\scalefont{1.10}#1}}
\usepackage[scaled=0.92]{PTSans}

\usepackage[usenames,dvipsnames]{xcolor}

\usepackage[parfill]{parskip}
\usepackage{afterpage}
\usepackage{framed}
\usepackage{nicefrac}
\usepackage{bm}
\usepackage{fixmath}

\usepackage[colorinlistoftodos,
           prependcaption,
           textsize=small,
           backgroundcolor=yellow,
           linecolor=lightgray,
           bordercolor=lightgray]{todonotes}

\usepackage{lineno}

\usepackage{ragged2e}

\DeclareRobustCommand{\parhead}[1]{\noindent\textbf{#1}~}

\usepackage{paralist}

\usepackage{graphicx}
\usepackage[labelfont=it,labelsep=period]{caption}
\usepackage{subfig}

\usepackage{booktabs}
\usepackage{arydshln} \usepackage{multirow}

\usepackage{natbib}

\usepackage{algorithm}
\usepackage{algorithmic}
\usepackage{listings}
\usepackage{fancyvrb}
\fvset{fontsize=\normalsize}

\usepackage[colorlinks,linktoc=all]{hyperref}
\usepackage[all]{hypcap}
\usepackage{url}

\usepackage[acronym,smallcaps,nowarn]{glossaries}

\usepackage{listings}
\lstdefinestyle{alp_style}{
    commentstyle=\color{OliveGreen},
    numberstyle=\tiny\color{black!60},
    stringstyle=\color{BrickRed},
    basicstyle=\ttfamily\scriptsize,
    breakatwhitespace=false,
    breaklines=true,
    captionpos=b,
    keepspaces=true,
    numbers=none,
    numbersep=5pt,
    showspaces=false,
    showstringspaces=false,
    showtabs=false,
    tabsize=2
}
\usepackage{lipsum}

\DeclareRobustCommand{\E}[2]{\mathbb{E}_{#1}\left[#2\right]}

\newcommand{\g}{\, | \,}
\newcommand{\prm}{\, ; \,}

\newcommand{\Lcal}{\mathcal{L}}
\newcommand{\Ncal}{\mathcal{N}}

\newcommand{\bzero}{\mathbf{0}}

\newcommand{\bx}{\mathbf{x}}
\newcommand{\bw}{\mathbf{w}}

\newcommand{\bI}{\mathbf{I}}

\newcommand{\cL}{\mathcal{L}}

\newcommand{\dd}{\, \mathrm{d}}
 \newacronym{ADVI}{advi}{automatic differentiation variational inference}
\newacronym{AR}{a{\small\&}r}{augment and reduce}

\newacronym{BBVI}{bbvi}{black-box variational inference}

\newacronym{CBOW}{cbow}{continuous bag-of-words}
\newacronym{CDF}{cdf}{cumulative distribution function}
\newacronym{CS-EFE}{cs-efe}{context selection for exponential family embeddings}
\newacronym{CTM}{ctm}{correlated topic model}

\newacronym[\glslongpluralkey={deep exponential families}]{DEF}{def}{deep exponential family}
\newacronym{DMIS}{dmis}{deterministic multiple importance sampling}

\newacronym{EFE}{efe}{exponential family embeddings}
\newacronym{ELBO}{elbo}{evidence lower bound}
\newacronym{EM}{em}{expectation maximization}
\newacronym{ETM}{etm}{\emph{embedded topic model}}
\newacronym{DETM}{detm}{\emph{deep embedded topic model}}

\newacronym{GNTS}{gn-ts}{gamma-normal time series model}
\newacronym{G-REP}{g-rep}{generalized reparameterization}

\newacronym{HMC}{hmc}{{H}amiltonian {M}onte {C}arlo}

\newacronym{KL}{kl}{{K}ullback-{L}eibler}

\newacronym{LDA}{lda}{latent {D}irichlet allocation}

\newacronym{MAP}{map}{\emph{maximum a posteriori}}
\newacronym{MCMC}{mcmc}{{M}arkov chain {M}onte {C}arlo}
\newacronym{MF}{mf}{matrix factorization}
\newacronym{MIS}{mis}{multiple importance sampling}

\newacronym{NVDM}{nvdm}{neural variational document model}

\newacronym{OBBVI}{o-bbvi}{overdispersed black-box variational inference}
\newacronym{OVE}{ove}{one-vs-each}

\newacronym{SIVI}{sivi}{semi-implicit variational inference}
\newacronym{SVI}{svi}{stochastic variational inference}

\newacronym{TMES}{tmes}{topic model in embedding space}

\newacronym{USIVI}{uivi}{unbiased implicit variational inference}

\newacronym{VAE}{vae}{variational autoencoder}
\newacronym{VEM}{vem}{variational expectation maximization}
\newacronym{VI}{vi}{variational inference}

\usepackage{times,latexsym}
\usepackage{url}
\usepackage[T1]{fontenc}

\usepackage{caption}

\usepackage[acceptedWithA]{tacl2018v2}

\hypersetup{citecolor=Violet}
\hypersetup{linkcolor=black}
\hypersetup{urlcolor=MidnightBlue}

\title{Topic Modeling in Embedding Spaces}
\author{
 Adji B. Dieng\\
 Columbia University \\
  {\sf abd2141@columbia.edu} \\
 \And
 Francisco J. R. Ruiz\\
  Columbia University\\Cambridge University\\
  {\sf f.ruiz@columbia.edu} \\
 \And
David M. Blei\\
 Columbia University \\
   {\sf david.blei@columbia.edu} 
}
\date{}

\begin{document}
\maketitle
\begin{abstract}
  Topic modeling analyzes documents to learn meaningful patterns of
  words.  However, existing topic models fail to learn interpretable
  topics when working with large and heavy-tailed vocabularies.  To
  this end, we develop the \gls{ETM}, a generative model of documents
  that marries traditional topic models with word embeddings.  In
  particular, it models each word with a categorical distribution whose natural
  parameter is the inner product between a word embedding and an embedding
  of its assigned topic.  To fit the \gls{ETM}, we develop an
  efficient amortized variational inference algorithm.  The \gls{ETM}
  discovers interpretable topics even with large vocabularies that
  include rare words and stop words.  It outperforms existing document
  models, such as \acrlong{LDA}, in terms of both topic quality and
  predictive performance.\footnote{Code for this work can be found at \url{https://github.com/adjidieng/ETM}.}
\end{abstract}

\section{Introduction}
\label{sec:introduction}
\glsresetall

Topic models are statistical tools for discovering the hidden semantic
structure in a collection of documents
\citep{blei2003latent,blei2012probabilistic}. Topic models and their
extensions have been applied to many fields, such as marketing,
sociology, political science, and the digital humanities.
\citet{boydgraber2017applications} provide a review.

Most topic models build on \gls{LDA} \citep{blei2003latent}.
\gls{LDA} is a hierarchical probabilistic model that represents each
topic as a distribution over terms and represents each document as a
mixture of the topics. When fit to a collection of documents, the
topics summarize their contents, and the topic proportions provide a
low-dimensional representation of each one.  \gls{LDA} can be fit to
large datasets of text by using variational inference and stochastic
optimization~\citep{Hoffman2010, Hoffman2013}.

\gls{LDA} is a powerful model and it is widely used.  However, it
suffers from a pervasive technical problem---it fails in the face of
large vocabularies. Practitioners must severely prune their
vocabularies in order to fit good topic models, i.e., those that are
both predictive and interpretable.
This is typically done by removing the most and least
frequent words.
On large collections, this pruning
may remove important terms and limit the scope of the models.  The
problem of topic modeling with large vocabularies has yet to be
addressed in the research literature.

In parallel with topic modeling came the idea of word embeddings.
Research in word embeddings begins with the neural language model of
\citet{Bengio:2003}, published in the same year and journal as
\citet{blei2003latent}. Word embeddings eschew the ``one-hot''
representation of words---a vocabulary-length vector of zeros with a
single one---to learn a distributed representation, one where words
with similar meanings are close in a lower-dimensional vector
space~\citep{Rumelhart:1973,bengio2006neural}.
As for topic models, researchers scaled
up embedding methods to large datasets
\citep{mikolov2013efficient,mikolov2013distributed,pennington2014glove,levy2014neural,mnih2013learning}.
Word embeddings have been extended and developed in many ways.  They
have become crucial in many applications of natural language
processing \citep{Li2017}, and they have also been extended to datasets
beyond text \citep{Rudolph2016}.

In this paper, we develop the \gls{ETM}, a topic model for word
embeddings. The \gls{ETM} enjoys the good properties of topic models
and the good properties of word embeddings.  As a topic model, it
discovers an interpretable latent semantic structure of the texts; as
a word embedding, it provides a low-dimensional representation of the
meaning of words.  It robustly accommodates large vocabularies and the
long tail of language data.

\begin{figure}[t]
	\centering
	\includegraphics[width=0.5\textwidth]{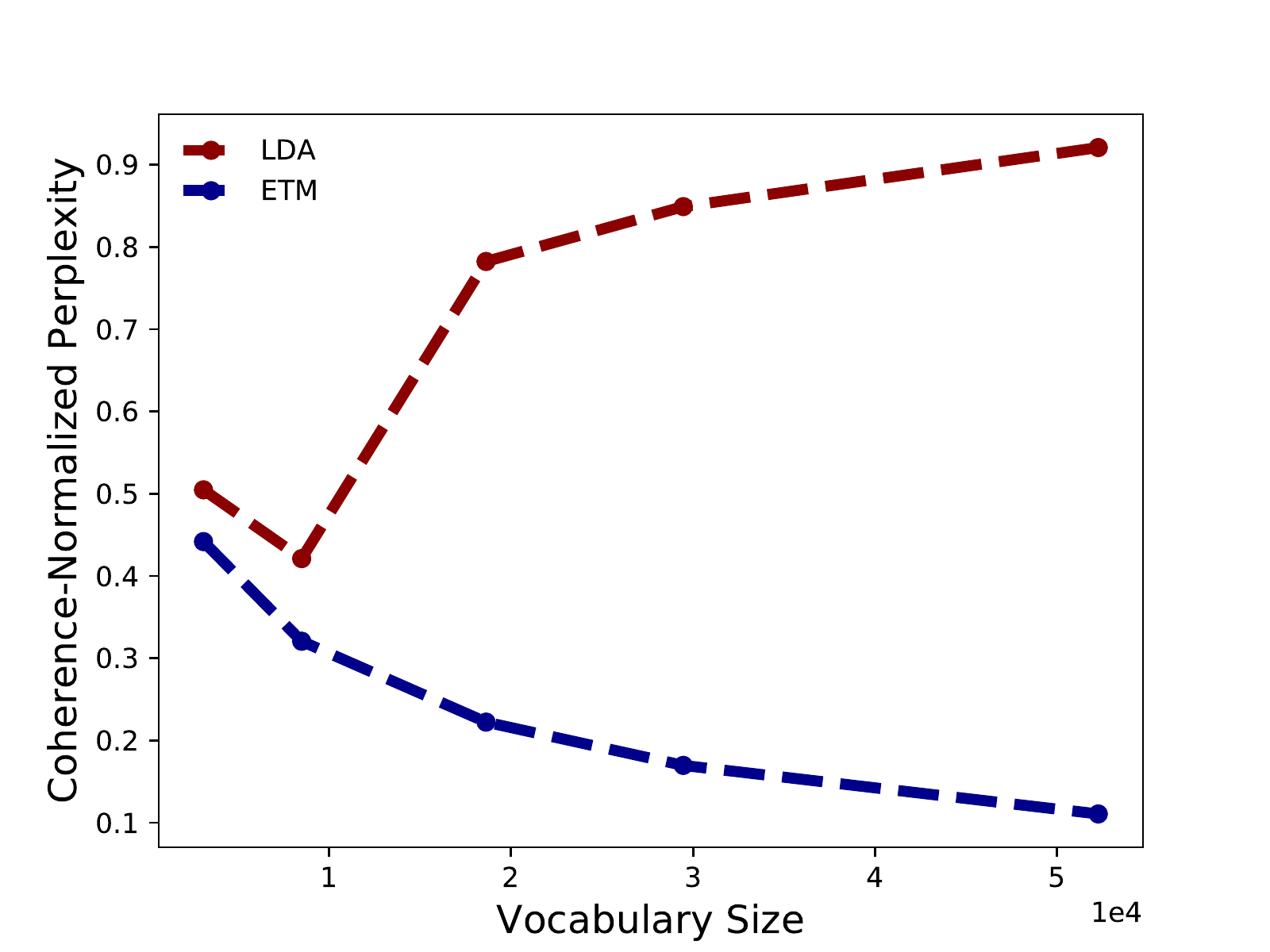}
	\caption{Ratio of the normalized held-out perplexity for document completion
	and the topic coherence as as a function of the vocabulary size for the \acrshort{ETM} and \acrshort{LDA}. While the performance of \gls{LDA} deteriorates for large vocabularies, the \gls{ETM} maintains good performance.
	\label{fig:log_lik_intro}}
\end{figure}

Figure\nobreakspace \ref {fig:log_lik_intro} illustrates the advantages.  This figure plots the
ratio between the predictive perplexity of held-out documents and the topic
coherence, as a function of the size of the vocabulary.
(The perplexity has been normalized by the vocabulary size.)
This is for a corpus of $11.2$K articles from the \textit{20NewsGroup} and for $100$ topics.  
The red line is \gls{LDA}; its performance deteriorates as the vocabulary size
increases---the predictive performance and the quality of the topics get worse. 
The blue line is the \gls{ETM};
it maintains good performance, even as the vocabulary size gets large.

\begin{figure*}[t]
\centering
\begin{minipage}{.45\textwidth}
  \centering
  \includegraphics[width=.8\linewidth]{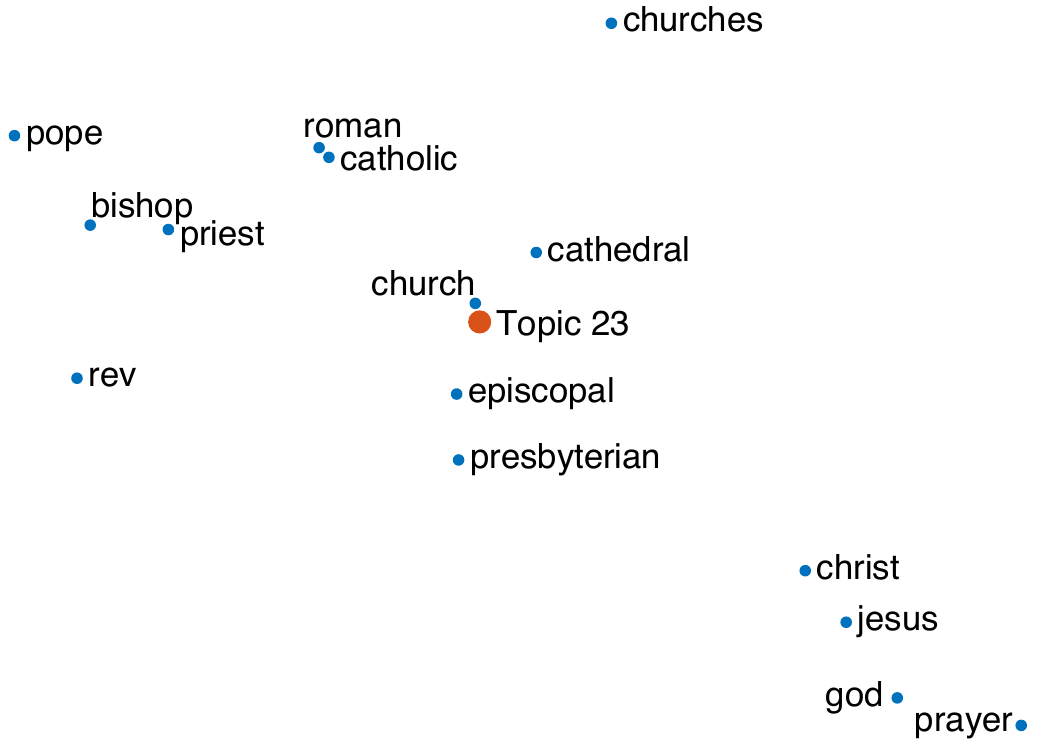}
  \captionof{figure}{A topic about Christianity found by the \acrshort{ETM} on \emph{The New York Times}. The topic is a point in the word embedding space.}
  \label{fig:topic_embedding_intro1}
\end{minipage}\hspace*{1cm}
\begin{minipage}{.45\textwidth}
  \centering
  \includegraphics[width=0.91\linewidth]{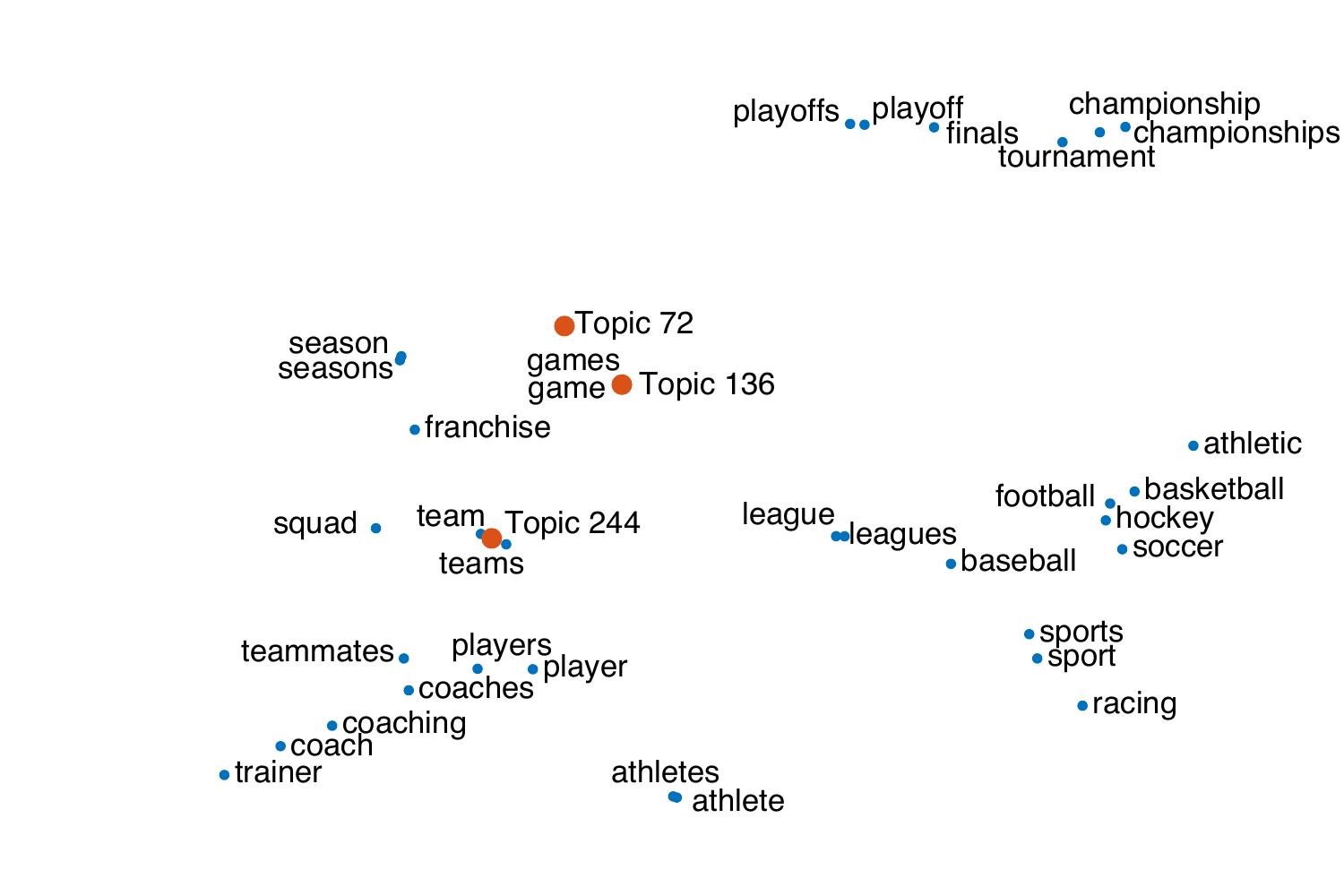}
  \captionof{figure}{Topics about sports found by the \acrshort{ETM}. Each topic is a point in the word embedding space.}
  \label{fig:topic_embedding_intro2}
\end{minipage}
\end{figure*}

Like \gls{LDA}, the \gls{ETM} is a generative probabilistic model:
each document is a mixture of topics and each observed word is
assigned to a particular topic.  In contrast to \gls{LDA},
the per-topic conditional probability of a term has a log-linear form
that involves a low-dimensional representation of the vocabulary.
Each term is represented by an embedding; each topic is a point in
that embedding space; and the topic's distribution over terms is
proportional to the exponentiated inner product of the topic's
embedding and each term's embedding.  Figures\nobreakspace \ref {fig:topic_embedding_intro1} and\nobreakspace  \ref {fig:topic_embedding_intro2}
show topics from a $300$-topic \gls{ETM} of \textit{The New York Times}. The figures show each topic's embedding and its closest words; these topics are about Christianity and sports.

Due to the topic representation in terms of a point in the embedding space,
the \gls{ETM} is also robust to the presence of stop words, unlike most common
topic models. When stop words are included in the vocabulary, the \gls{ETM}
assigns topics to the corresponding area of the embedding space
(we demonstrate this in Section\nobreakspace \ref {sec:experiments}).

As for most topic models, the posterior of the topic
proportions is intractable to compute.  We derive an efficient
algorithm for approximating the posterior with variational
inference~\citep{Jordan1999,Hoffman2013,Blei2017} and additionally use
amortized inference to efficiently approximate the topic
proportions~\citep{kingma2014autoencoding, rezende2014stochastic}. The
resulting algorithm fits the \gls{ETM} to large corpora with large
vocabularies.  The algorithm for the \gls{ETM} can either use previously fitted
word embeddings, or fit them jointly with the rest of parameters.
(In particular, Figures\nobreakspace  \ref {fig:log_lik_intro} to\nobreakspace  \ref {fig:topic_embedding_intro2} 
were obtained with the version of the \gls{ETM} that obtains pre-fitted
skip-gram word embeddings.)

We compared the performance of the \gls{ETM} to \gls{LDA} and the
\gls{NVDM}, a form of multinomial matrix factorization.  The \gls{ETM}
provides good predictive performance, as measured by held-out
log-likelihood on a document completion task
\citep{wallach2009evaluation}. It also provides meaningful topics, as
measured by topic coherence \citep{mimno2011optimizing} and topic
diversity, a metric that also indicates the quality of the topics.
The \gls{ETM} is especially robust to large vocabularies.

 \section{Related Work}
\label{sec:related}

This work develops a new topic model that extends \gls{LDA}.
\gls{LDA} has been extended in many ways, and topic modeling has
become a subfield of its own.  For a review, see
\citet{blei2012probabilistic} and \citet{boydgraber2017applications}.

One of the goals in developing the \gls{ETM} is to incorporate word
similarity into the topic model, and there is previous research that
shares this goal.  These methods either modify the topic
priors~\citep{petterson2010word, zhao2017metalda, shi2017jointly,
  zhao2017word} or the topic assignment
priors~\citep{xie2015incorporating}. For example
\citet{petterson2010word} use a word similarity graph (as given by a
thesaurus) to bias \gls{LDA} towards assigning similar words to
similar topics. As another example, \citet{xie2015incorporating} model
the per-word topic assignments of \gls{LDA} using a Markov random
field to account for both the topic proportions and the topic
assignments of similar words.  These methods use word
similarity as a type of ``side information'' about language; in
contrast, the \gls{ETM} directly models the similarity (via
embeddings) in its generative process of words.

Other work has extended \gls{LDA} to directly involve word embeddings.
One common strategy is to convert the discrete text into continuous
observations of embeddings, and then adapt \gls{LDA} to generate
real-valued data~\citep{das2015gaussian, xun2016topic,
  batmanghelich2016nonparametric, xun2017correlated}.  With this
strategy, topics are Gaussian distributions with latent means and
covariances, and the likelihood over the embeddings is modeled with a
Gaussian~\citep{das2015gaussian} or a Von-Mises Fisher
distribution~\citep{batmanghelich2016nonparametric}.  The \gls{ETM}
differs from these approaches in that it is a model of categorical
data, one that goes through the embeddings matrix.  Thus it does not
require pre-fitted embeddings and, indeed, can learn embeddings as
part of its inference process.

There have been a few other ways of combining \gls{LDA} and
embeddings. \citet{nguyen2015improving} mix the likelihood defined by
\gls{LDA} with a log-linear model that uses pre-fitted word
embeddings; \citet{bunk2018welda} randomly replace words drawn from a
topic with their embeddings drawn from a Gaussian; and
\citet{xu2018distilled} adopt a geometric perspective, using
Wasserstein distances to learn topics and word embeddings jointly.

Another thread of recent research improves topic modeling inference
through deep neural networks \citep{srivastava2017autoencoding,
  card2017neural,cong2017deep, zhang2018whai}. Specifically, these
methods reduce the dimension of the text data through amortized
inference and the variational
auto-encoder~\citep{kingma2014autoencoding, rezende2014stochastic}.
To perform inference in the \gls{ETM}, we also avail ourselves of
amortized inference methods \citep{Gershman2014}.

Finally, as a document model, the \gls{ETM} also relates to works that
learn per-document representations as part of an embedding
model~\citep{Le2014, moody2016mixing, miao2016neural}. In contrast to
these works, the document variables in the \gls{ETM} are part of a larger
probabilistic topic model.

 \section{Background}
\label{sec:background}

The \acrshort{ETM} builds on two main ideas, \gls{LDA} and word
embeddings. Consider a corpus of $D$ documents, where the vocabulary
contains $V$ distinct terms. Let $w_{dn}\in\{1,\ldots,V \}$ denote the
$n^{\textrm{th}}$ word in the $d^{\textrm{th}}$ document.

\parhead{Latent Dirichlet allocation.}  \gls{LDA} is a probabilistic
generative model of documents \citep{blei2003latent}. It posits $K$
topics $\beta_{1:K}$, each of which is a distribution over the
vocabulary.  \gls{LDA} assumes each document comes from a mixture of
topics, where the topics are shared across the corpus and the mixture
proportions are unique for each document.  The generative process for
each document is the following:
\begin{compactenum}
\item Draw topic proportion $\theta_d \sim \textrm{Dirichlet}(\alpha_{\theta})$.
\item For each word $n$ in the document:
  \begin{compactenum}
    \setlength{\itemindent}{-0.3cm}
  \item Draw topic assignment $z_{dn} \sim \text{Cat}(\theta_d)$.
  \item Draw word $w_{dn} \sim \text{Cat}(\beta_{z_{dn}})$.
  \end{compactenum}
\end{compactenum}
Here, Cat$(\cdot)$ denotes the categorical distribution. \gls{LDA}
places a Dirichlet prior on the topics,
$\beta_k\sim \textrm{Dirichlet}(\alpha_{\beta})$ for $k=1,\ldots,K$.
The concentration parameters $\alpha_{\beta}$ and $\alpha_{\theta}$ of the
Dirichlet distributions are fixed model hyperparameters.

\parhead{Word embeddings.}  Word embeddings provide models of language
that use vector representations of
words~\citep{Rumelhart:1973,Bengio:2003}. The word representations are
fitted to relate to meaning, in that words with similar meanings will
have representations that are close.  (In embeddings, the ``meaning''
of a word comes from the contexts in which it is used.)

We focus on the \gls{CBOW} variant of word embeddings
\citep{mikolov2013distributed}. In \gls{CBOW}, the likelihood of each
word $w_{dn}$ is
\begin{equation}\label{eq:cbow}
  w_{dn} \sim \textrm{softmax}(\rho^\top \alpha_{dn}).
\end{equation}
The embedding matrix $\rho$ is a $L\times V$ matrix whose columns
contain the embedding representations of the vocabulary,
$\rho_v\in\mathbb{R}^L$. The vector $\alpha_{dn}$ is the \emph{context
  embedding}. The context embedding is the sum of the context
embedding vectors ($\alpha_v$ for each word $v$) of the words surrounding
$w_{dn}$.

\section{The Embedded Topic Model}
\label{sec:model}

The \gls{ETM} is a topic model that uses embedding representations of
both words and topics.  It contains two notions of latent
dimension. First, it embeds the vocabulary in an $L$-dimensional
space.  These embeddings are similar in spirit to classical word
embeddings.  Second, it represents each document in terms of $K$
latent topics.

In traditional topic modeling, each topic is a full distribution over
the vocabulary. In the \gls{ETM}, however, the $k^{\textrm{th}}$ topic is a vector
$\alpha_k\in\mathbb{R}^L$ in the embedding space. We call $\alpha_k$ a
\emph{topic embedding}---it is a distributed representation of the
$k^{\textrm{th}}$ topic in the semantic space of words.

In its generative process, the \gls{ETM} uses the topic embedding to
form a per-topic distribution over the vocabulary. Specifically, the
\gls{ETM} uses a log-linear model that takes the inner product of the
word embedding matrix and the topic embedding.  With this form, the
\gls{ETM} assigns high probability to a word $v$ in topic $k$ by
measuring the agreement between the word's embedding and the topic's
embedding.

Denote the $L \times V$ word embedding matrix by $\rho$; the column
$\rho_v$ is the embedding of $v$.  Under the \gls{ETM}, the generative
process of the $d^{\textrm{th}}$ document is the following:
\begin{compactitem}
\item[1.] Draw topic proportions $\theta_d \sim \mathcal{LN}(0,I).$
\item[2.] For each word $n$ in the document:
  \begin{compactitem}
    \setlength{\itemindent}{-0.3cm}
  \item[a.] Draw topic assignment $z_{dn} \sim \text{Cat}(\theta_d).$
  \item[b.] Draw the word $w_{dn} \sim \text{softmax}(\rho^\top\alpha_{z_{dn}})$.
  \end{compactitem}
\end{compactitem}
In Step 1, $\mathcal{LN}(\cdot)$ denotes the logistic-normal
distribution~\citep{Aitchison:1980,blei2007correlated}; it transforms
a standard Gaussian random variable to the simplex. A draw $\theta_d$
from this distribution is obtained as
\begin{align}
  \label{eq:logistic-normal}
  \delta_d \sim \Ncal\left(0,I\right); \quad  \theta_d =
  \text{softmax}(\delta_d).
\end{align}
(We replaced the Dirichlet with the logistic normal to more easily use
reparameterization in the inference algorithm; see
Section\nobreakspace \ref {sec:inference}.)

Steps 1 and 2a are standard for topic modeling: they represent
documents as distributions over topics and draw a topic assignment for
each observed word. Step 2b is different; it uses the embeddings of
the vocabulary $\rho$ and the assigned topic embedding
$\alpha_{z_{dn}}$ to draw the observed word from the assigned topic, as
given by $z_{dn}$.

The topic distribution in Step 2b mirrors the \gls{CBOW} likelihood in
Eq.\nobreakspace \ref {eq:cbow}. Recall \gls{CBOW} uses the surrounding words to form
the context vector $\alpha_{dn}$.  In contrast, the \gls{ETM} uses the
topic embedding $\alpha_{z_{dn}}$ as the context vector, where the
assigned topic $z_{dn}$ is drawn from the per-document variable
$\theta_d$.  The \gls{ETM} draws its words from a document context,
rather than from a window of surrounding words.

The \gls{ETM} likelihood uses a matrix of word embeddings $\rho$, a
representation of the vocabulary in a lower dimensional space.  In
practice, it can either rely on previously fitted embeddings or learn
them as part of its overall fitting procedure.  When the \gls{ETM}
learns the embeddings as part of the fitting procedure, it
simultaneously finds topics and an embedding space.

When the \gls{ETM} uses previously fitted embeddings, it learns the
topics of a corpus in a particular embedding space.  This strategy is
particularly
useful when there are words in the embedding that are not used in the
corpus.  The \gls{ETM} can hypothesize how those words fit in to the
topics because it can calculate $\rho_v^\top \alpha_k$, even for words
$v$ that do not appear in the corpus.

\section{Inference and Estimation}
\label{sec:inference}

We are given a corpus of documents $\{\bw_{1}, \ldots, \bw_{D}\}$,
where $\bw_d$ is a collection of $N_d$ words.  How do we fit the
\gls{ETM}?

\parhead{The marginal likelihood.} The parameters of the \gls{ETM} are
the embeddings $\rho_{1:V}$ and the topic embeddings
$\alpha_{1:K}$; each $\alpha_k$ is a point
in the embedding space.  We maximize the marginal likelihood of the
documents,
\begin{align}
  \label{eq:marginal}
  \cL(\alpha, \rho) = \sum_{d=1}^{D} \log p(\bw_d \g \alpha, \rho).
\end{align}

The problem is that the marginal likelihood of each document is
intractable to compute.  It involves a difficult integral over the
topic proportions, which we write in terms of the untransformed
proportions $\delta_d$ in Eq.\nobreakspace \ref {eq:logistic-normal},
\begin{align}
  \label{eq:integral}
  p(\bw_d \g \alpha, \rho) =
  \int p(\delta_d)
  \prod_{n=1}^{N_d}
  p(w_{dn} \g \delta_d, \alpha, \rho) \dd \delta_d.
\end{align}
The conditional distribution of each word marginalizes out the topic
assignment $z_{dn}$,
\begin{align}
  \label{eq:likelihood}
  p(w_{dn} \g \delta_d, \alpha, \rho)
  &= \sum_{k=1}^{K} \theta_{dk} \beta_{k,w_{dn}}.
\end{align}
Here, $\theta_{dk}$ denotes the (transformed) topic proportions
(Eq.\nobreakspace \ref {eq:logistic-normal}) and $\beta_{kv}$ denotes a traditional
``topic,'' i.e., a distribution over words, induced by the word embeddings
$\rho$ and the topic embedding $\alpha_k$,
\begin{align}
  \label{eq:topic}
  \beta_{kv} = \textrm{softmax}(\rho^\top \alpha_k)\big|_v.
\end{align}
Eqs.\nobreakspace  \ref {eq:integral} to\nobreakspace  \ref {eq:topic}  flesh out the likelihood in
Eq.\nobreakspace \ref {eq:marginal}.

\parhead{Variational inference.} We sidestep the intractable integral
with variational inference \citep{Jordan1999,Blei2017}. Variational
inference optimizes a sum of per-document bounds on the log of the
marginal likelihood of Eq.\nobreakspace \ref {eq:integral}.  There are two sets of
parameters to optimize: the model parameters, as described above, and
the variational parameters, which tighten the bounds on the marginal
likelihoods.

To begin, posit a family of distributions of the untransformed topic
proportions $q(\delta_d\prm \bw_d, \nu)$.  We use amortized
inference, where the variational distribution of $\delta_d$ depends on
both the document $\bw_d$ and shared variational parameters $\nu$.  In
particular $q(\delta_d\prm \bw_d, \nu)$ is a Gaussian whose mean and
variance come from an ``inference network,'' a neural network
parameterized by $\nu$ \citep{kingma2014autoencoding}.
The inference network ingests the document $\bw_d$ and outputs a
mean and variance of $\delta_d$. (To accommodate documents of varying
length, we form the input of the inference network by normalizing the
bag-of-word representation of the document by the number of words $N_d$.)

We use this family of variational distributions to bound the log-marginal
likelihood.  The \gls{ELBO} is a function of the model
parameters and the variational parameters,
\begin{align}
  \label{eq:elbo}
  \cL (\alpha, \rho, \nu) & =
        \sum_{d=1}^{D}
        \sum_{n=1}^{N_d}
        \E{q}{\log p(w_{nd} \g \delta_d, \rho, \alpha)} \nonumber\\
       & - \sum_{d=1}^{D}
        \mathrm{KL}(q(\delta_d ; \bw_d, \nu) \; || \; p(\delta_d)).
\end{align}
As a function of the variational parameters, the first term encourages
them to place mass on topic proportions $\delta_d$ that explain the
observed words; the second term encourages them to be close to the
prior $p(\delta_d)$.  As a function of the model parameters, this
objective maximizes the expected complete log-likelihood,
$\sum_{d} \log p(\delta_d, \bw_d \g \alpha, \rho)$.

We optimize the \gls{ELBO} with respect to both the model parameters
and the variational parameters. We use stochastic optimization,
forming noisy gradients by taking Monte Carlo approximations of the
full gradient through the reparameterization trick
\citep{kingma2014autoencoding,Titsias2014_doubly,rezende2014stochastic}.
We also use data subsampling to handle large collections of documents
\citep{Hoffman2013}.  We set the learning rate with Adam \citep{Kingma2015}.
The procedure is shown in Algorithm\nobreakspace \ref {alg:etm}, where the notation
$\textrm{NN}(\mathbf{x}\prm \nu)$ represents a neural network with input
$\mathbf{x}$ and parameters $\nu$.

\begin{algorithm}[t]
  \caption{Topic modeling with the \gls{ETM}}
  \label{alg:etm}
  \small
  \begin{algorithmic}
    \STATE Initialize model and variational parameters
    \FOR{iteration $i = 1, 2, \ldots$}
    \STATE Compute $\beta_k = \text{softmax}(\rho^\top \alpha_k)$ for each topic $k$
    \STATE Choose a minibatch $\mathcal{B}$ of documents
    \FOR{each document $d$ in $\mathcal{B}$}
    \STATE Get normalized bag-of-word representat.\ $\bx_d$
    \STATE Compute $\mu_d = \textrm{NN}(\bx_d\prm \nu_{\mu})$
    \STATE Compute $\Sigma_d = \textrm{NN}(\bx_d\prm \nu_{\Sigma})$
                \STATE Sample $\theta_d \sim \Lcal\Ncal(\mu_d, \Sigma_d)$
    \FOR{each word in the document}
    \STATE Compute $p(w_{dn} \g \theta_d) = \theta_d^\top\beta_{\cdot,w_{dn}}$
    \ENDFOR
    \ENDFOR
    \STATE Estimate the \acrshort{ELBO} and its gradient (backprop.)
    \STATE Update model parameters $\alpha_{1:K}$
    \STATE Update variational parameters ($\nu_\mu$, $\nu_\Sigma$)
    \ENDFOR
  \end{algorithmic}
\end{algorithm}

\section{Empirical Study}
\label{sec:experiments}

\begin{table*}[t]
  \centering \captionof{table}{Word embeddings learned by all document
    models (and skip-gram) on the \emph{New York Times} with
    vocabulary size $118{,}363$.}  \vskip 0.1in
  \makebox[\textwidth][c]{
  \begin{tabular}{llllllll}
\multicolumn{4}{c}{Skip-gram embeddings} & \multicolumn{4}{c}{\gls{ETM} embeddings} \\
\cmidrule(r){1-4} \cmidrule(r){5-8}
\bf{love} & \bf{family} & \bf{woman} & \bf{politics}  & \bf{love} &  \bf{family} & \bf{woman} & \bf{politics}\\
loved & families  & man & political & joy & children & girl & political\\
passion &grandparents & girl & religion & loves & son & boy & politician \\
loves & mother & boy & politicking & loved & mother &  mother & ideology \\
affection & friends & teenager & ideology & passion & father & daughter  & speeches\\
adore & relatives  & person & partisanship & wonderful & wife & pregnant & ideological\\
\cmidrule(r){1-4} \cmidrule(r){5-8}
 \end{tabular}
 }
   \vskip 0.1in
   \makebox[\textwidth][c]{
  \begin{tabular}{llllllll}
\multicolumn{4}{c}{\gls{NVDM} embeddings} & \multicolumn{4}{c}{$\Delta$-\gls{NVDM} embeddings} \\
\cmidrule(r){1-4} \cmidrule(r){5-8}
\bf{love} & \bf{family} & \bf{woman} & \bf{politics}  & \bf{love} &  \bf{family} & \bf{woman} & \bf{politics}\\
loves & sons & girl & political & miss & home & life &  political   \\
passion & life & women & politician & young & father & marriage &  faith  \\
wonderful & brother & man & politicians & born & son &  women & marriage   \\
joy & son & pregnant & politically & dream & day &  read &  politicians  \\
beautiful &  lived & boyfriend & democratic & younger & mrs &  young & election   \\
\cmidrule(r){1-4} \cmidrule(r){5-8}
 \end{tabular}
 }
\label{tab:embeddings}
\end{table*}

\begin{table*}[t]
  \centering \small \captionof{table}{Top five words of seven most
    used topics from different document models on $1.8$M documents of
    the \emph{New York Times} corpus with vocabulary size $212{,}237$
    and $K=300$ topics.}  \vskip 0.1in
 \begin{tabular}{llllllll}
 \toprule
\multicolumn{7}{c}{LDA}\\
 \hline
     time & year & officials & mr & city & percent & state  \\
     day & million & public & president & building & million & republican  \\
     back & money & department & bush & street  & company  & party  \\
     good & pay & report  & white & park & year  & bill \\
     long & tax & state  & clinton & house & billion  & mr  \\
     \midrule 
\multicolumn{7}{c}{\gls{NVDM}}\\
\hline
     scholars & japan & gansler & spratt & assn & ridership & pryce \\
     gingrich & tokyo & wellstone & tabitha & assoc & mtv & mickens  \\
     funds & pacific & mccain & mccorkle & qtr & straphangers & mckechnie   \\
     institutions & europe & shalikashvili & cheetos & yr & freierman & mfume \\
     endowment & zealand & coached & vols  & nyse  & riders & filkins \\
\midrule
\multicolumn{7}{c}{$\Delta$-\gls{NVDM}}\\
\hline
     concerto & servings & nato & innings & treas & patients & democrats  \\
     solos & tablespoons & soviet & scored & yr & doctors  & republicans \\
     sonata & tablespoon & iraqi & inning & qtr & medicare & republican \\
     melodies & preheat  &  gorbachev & shutout & outst  & dr & senate \\
     soloist & minced &  arab & scoreless & telerate & physicians & dole \\
\midrule
\multicolumn{7}{c}{Labeled \gls{ETM}}\\
\hline
     music & republican & yankees & game &  wine &  court & company  \\
     dance & bush & game & points & restaurant &  judge & million \\
     songs & campaign & baseball & season & food &  case & stock \\
     opera & senator & season & team & dishes &  justice & shares \\
     concert & democrats & mets & play & restaurants &  trial & billion  \\
\midrule 
\multicolumn{7}{c}{\gls{ETM}}\\
\hline
     game & music & united & wine & company & yankees &  art\\
     team & mr & israel & food & stock & game & museum    \\
     season & dance & government & sauce &  million & baseball & show \\
     coach & opera & israeli & minutes & companies & mets &  work \\
     play &  band & mr &  restaurant & billion &  season & artist  \\
 \bottomrule
 \end{tabular}
\label{tab:topics}
\end{table*}

We study the performance of the \gls{ETM} and compare it to other
unsupervised document models.  A good document model should provide
both coherent patterns of language and an accurate distribution of
words, so we measure performance in terms of both predictive accuracy
and topic interpretability.  We measure accuracy with log-likelihood
on a document completion task \citep{rosenzvi2004author,wallach2009evaluation};
we measure topic interpretability as a
blend of topic coherence and diversity.  We find that, of the interpretable
models, the \gls{ETM} is the one that provides better predictions and
topics.

In a separate analysis (Section\nobreakspace \ref {subsec:experiments_stopwords}), we study
the robustness of each method in the presence of stop words. Standard
topic models fail in this regime---since stop words appear in many documents,
every learned topic includes some stop words, leading to poor topic
interpretability.
In contrast, the \gls{ETM} is able to use the information from the word
embeddings to provide interpretable topics.\footnote{Code is available upon request and will be released after publication.}

\parhead{Corpora.} We study the \emph{20Newsgroups} corpus and the
\emph{New York Times} corpus.

The \emph{20Newsgroup} corpus is a collection of newsgroup posts. We
preprocess the corpus by filtering stop words, words with document
frequency above 70\%, and tokenizing.  To form the vocabulary, we keep
all words that appear in more than a certain number of documents, and
we vary the threshold from 100 (a smaller vocabulary, where
$V = 3{,}102$) to 2 (a larger vocabulary, where $V=52{,}258$). After
preprocessing, we further remove one-word documents from the
validation and test sets.
We split the corpus into a training set of $11{,}260$ documents, a test
set of $7{,}532$ documents, and a validation set of $100$ documents.

The \emph{New York Times} corpus is a larger collection of news
articles.  It contains more than $1.8$ million articles, spanning the
years 1987--2007.  We follow the same preprocessing steps as for
\emph{20Newsgroups}.  We form versions of this corpus with
vocabularies ranging from $V=5{,}921$ to $V=212{,}237$.  After
preprocessing, we use $85\%$ of the documents for training, $10\%$ for
testing, and $5\%$ for validation.

\glsreset{LDA}
\glsreset{NVDM}

\parhead{Models.} We compare the performance of the \gls{ETM} with two
other document models: \gls{LDA} and the \gls{NVDM}.

\gls{LDA} \citep{blei2003latent} is a standard topic model that posits
Dirichlet priors for the topics $\beta_k$ and topic proportions
$\theta_d$. (We set the prior hyperparameters to $1$.) It is a
conditionally conjugate model, amenable to variational inference with
coordinate ascent.  We consider \gls{LDA} because it is the most
commonly used topic model, and it has a similar generative process as
the \gls{ETM}.

The \gls{NVDM} \citep{miao2016neural} is a multinomial factor model of
documents; it posits the likelihood
$w_{dn}\sim \text{softmax}(\beta^\top \theta_d)$, where the
$K$-dimensional vector $\theta_d\sim \mathcal{N}(\bzero, \bI_K)$ is a
per-document variable, and $\beta$ is a real-valued matrix of size
$K \times V$.  The \gls{NVDM} uses a per-document real-valued latent vector
$\theta_d$ to average over the embedding matrix $\beta$ in the logit
space.  Like the \gls{ETM}, the \gls{NVDM} uses amortized variational
inference to jointly learn the approximate posterior over the document
representation $\theta_d$ and the model parameter $\beta$.

\gls{NVDM} is not interpretable as a topic model; its latent variables
are unconstrained.  We study a more interpretable variant of the
\gls{NVDM} which constrains $\theta_d$ to lie in the simplex,
replacing its Gaussian prior with a logistic
normal~\citep{Aitchison:1980}.  (This can be thought of as a
semi-nonnegative matrix factorization.)  We call this document model
$\Delta$-\gls{NVDM}.

We study two variants of the \gls{ETM}, one where the word embeddings
are pre-fitted and one where they are learned jointly with the rest of
the parameters.  The variant with pre-fitted embeddings is called the
``labeled \gls{ETM}.'' We use skip-gram embeddings
\citep{mikolov2013distributed}.

\parhead{Algorithm settings.}  Given a corpus, each model comes with an
approximate posterior inference problem.  We use variational inference
for all of the models and employ \gls{SVI} \citep{Hoffman2013} to
speed up the optimization. The minibatch size is $1{,}000$ documents.
For \gls{LDA}, we set the learning rate as suggested by \citet{Hoffman2013}: the
delay is $10$ and the forgetting factor is $0.85$.

Within \gls{SVI}, \gls{LDA} enjoys coordinate ascent variational
updates, with $5$ inner steps to optimize the local variables.  For
the other models, we use amortized inference over the local variables
$\theta_d$.  We use $3$-layer inference networks and we set the local
learning rate to $0.002$. We use $\ell_2$ regularization on the
variational parameters (the weight decay parameter is
$1.2\times 10^{-6}$).

\begin{figure*}[t]
  \centering
  \subfloat[Topic quality as measured by normalized product of topic coherence and topic diversity (the higher the better) vs.\ predictive performance as measured by normalized log-likelihood on document completion (the higher the better) on the \emph{20NewsGroup} dataset.\label{fig:scatter_20ng}]{\includegraphics[width=\textwidth]{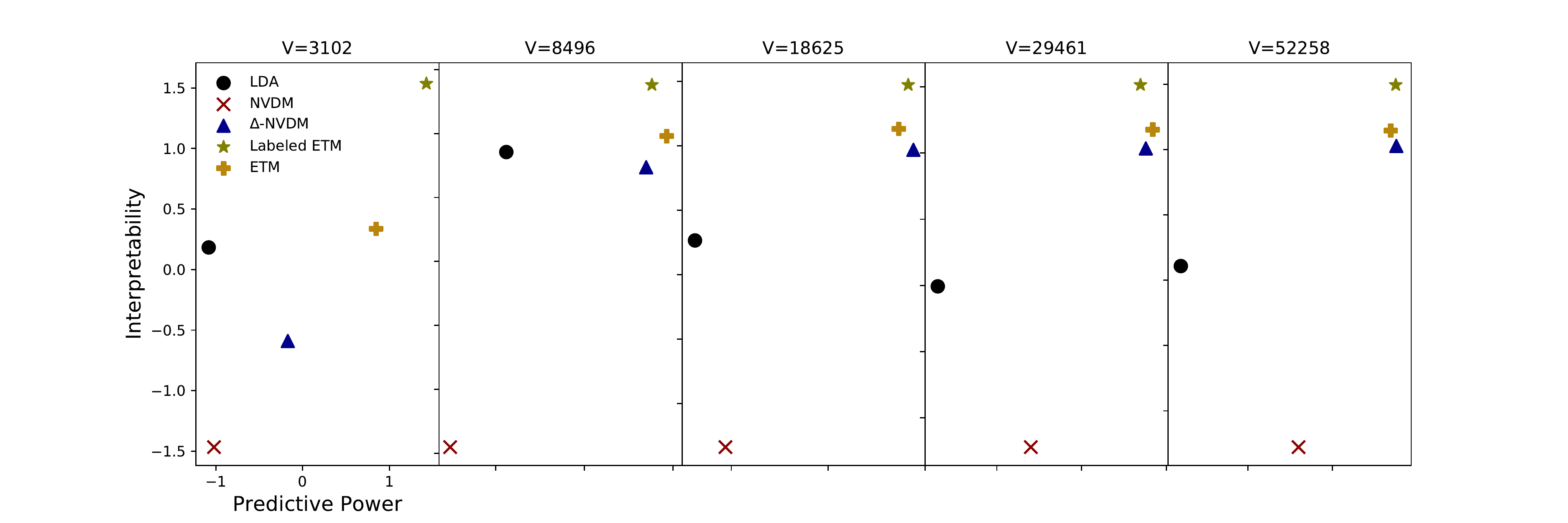}} \\
  \subfloat[Topic quality as measured by normalized product of topic coherence
  and topic diversity (the higher the better) vs.\ predictive performance as measured by normalized
  log-likelihood on document completion (the higher the better)
  on the \emph{New York Times}
  dataset.\label{fig:scatter_nyt}]{\includegraphics[width=\textwidth]{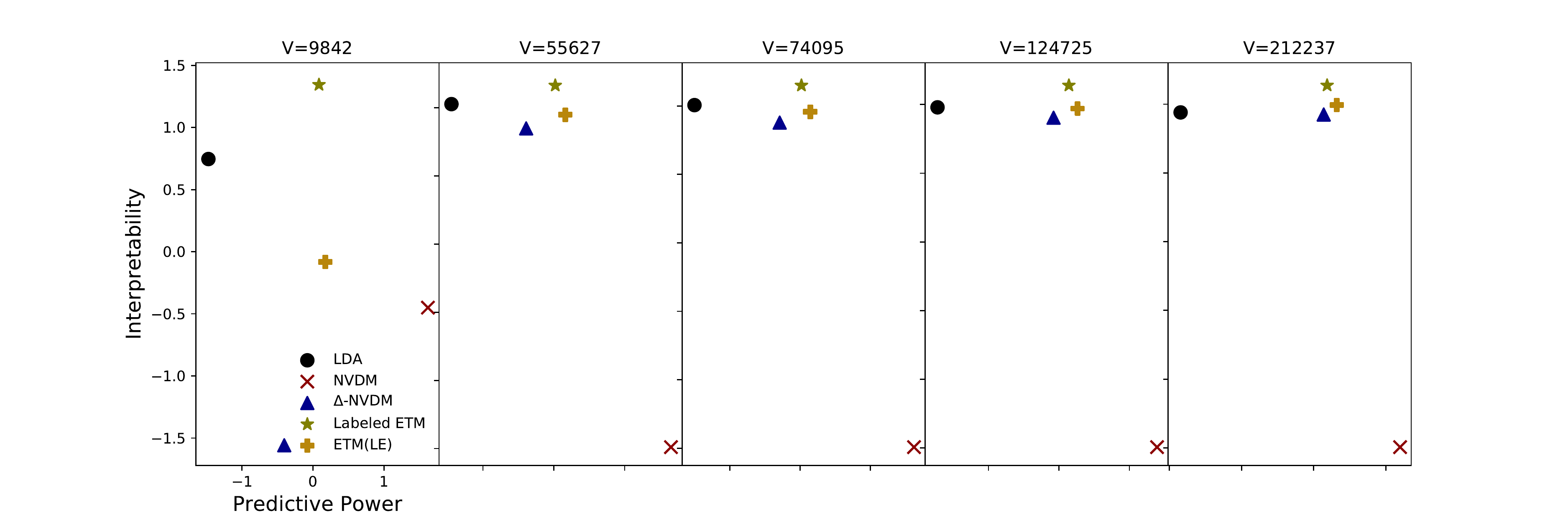}}
  \caption{Performance on the \emph{20NewsGroups} and the New York
    Times datasets for different vocabulary sizes. On both plots,
    better models are on the top right corner. Overall, the \gls{ETM}
    is a better topic model.}
  \label{fig:scatter}
\end{figure*}

\parhead{Qualitative results.}  We first examine the embeddings.  The
\gls{ETM}, \gls{NVDM}, and $\Delta$-\gls{NVDM} all involve a word
embedding.  We illustrate them by fixing a set of terms and
calculating the words that occur in the neighborhood around them.  For
comparison, we also illustrate word embeddings learned by the
skip-gram model.

Table\nobreakspace \ref {tab:embeddings} illustrates the embeddings of the different
models.  All the methods provide interpretable embeddings---words with
related meanings are close to each other. The \gls{ETM} and the
\gls{NVDM} learn embeddings that are similar to those from the
skip-gram.  The embeddings of $\Delta$-\gls{NVDM} are different; the
simplex constraint on the local variable changes the nature of the
embeddings.

We next look at the learned topics. Table\nobreakspace \ref {tab:topics} displays the $7$
most used topics for all methods, as given by the average of the topic
proportions $\theta_d$.  \gls{LDA} and the \gls{ETM} both provide
interpretable topics.  Neither \gls{NVDM} nor $\Delta$-\gls{NVDM}
provide interpretable topics; their model parameters $\beta$ 
are not interpretable as distributions over the vocabulary that mix to form documents.

\parhead{Quantitative results.} We next study the models
quantitatively.  We measure the quality of the topics and the
predictive performance of the model.  We found that among models with
interpretable topics, the \gls{ETM} provides the best predictions.

We measure topic quality by blending two metrics: topic coherence and
topic diversity.  Topic coherence is a quantitative measure of the
interpretability of a topic \citep{mimno2011optimizing}.  It is the average pointwise mutual
information of two words drawn randomly from the same document \citep{lau2014machine},
\begin{equation*}
  \textrm{TC} = \frac{1}{K}\sum_{k=1}^{K} \frac{1}{45} \sum_{i=1}^{10}
  \sum_{j=i+1}^{10} f(w_i^{(k)}, w_j^{(k)}),
\end{equation*}
where $\{w_1^{(k)},\ldots,w_{10}^{(k)}\}$ denotes the top-$10$ most
likely words in topic $k$.  Here, $f(\cdot,\cdot)$ is the normalized
pointwise mutual information,
\begin{equation*}
  f(w_i, w_j) = \frac{\log\frac{P(w_i,w_j)}{P(w_i)P(w_j)}}{-\log
    P(w_i,w_j)}.
\end{equation*}
The quantity $P(w_i,w_j)$ is the probability of words $w_i$ and $w_j$
co-occurring in a document and $P(w_i)$ is the marginal probability of
word $w_i$.  We approximate these probabilities with empirical counts.

The idea behind topic coherence is that a coherent topic will display
words that tend to occur in the same documents. In other words, the
most likely words in a coherent topic should have high mutual
information. Document models with higher topic coherence are more
interpretable topic models.

We combine coherence with a second metric, topic diversity.  We define
topic diversity to be the percentage of unique words in the top $25$
words of all topics. Diversity close to $0$ indicates redundant
topics; diversity close to $1$ indicates more varied topics.  We define the
overall metric for the quality of a model's topics as the product of
its topic diversity and topic coherence.

A good topic model also provides a good distribution of language.  To
measure predictive quality, we calculate log likelihood on a document
completion task~\citep{rosenzvi2004author,wallach2009evaluation}. We
divide each test document into two sets of words.  The first half is
observed: it induces a distribution over topics which, in turn,
induces a distribution over the next words in the document.  We then
evaluate the second half under this distribution. A good document
model should provide higher log-likelihood on the second half. (For
all methods, we approximate the likelihood by setting $\theta_d$ to
the variational mean.)

We study both corpora and with different
vocabularies. Figure\nobreakspace \ref {fig:scatter} shows topic quality as a function of
predictive power. (To ease visualization, we normalize both metrics by
subtracting the mean and dividing by the standard deviation.) The best
models are on the upper right corner.

\gls{LDA} predicts worst in almost all settings. On
\emph{20NewsGroups}, the \gls{NVDM}'s predictions are in general
better than \gls{LDA} but worse than for the other methods; on the
\emph{New York Times}, the \gls{NVDM} gives the best
predictions. However, topic quality for the \gls{NVDM} is far below
the other methods.  (It does not provide ``topics'', so we assess the
interpretability of its $\beta$ matrix.)  In prediction, both versions
of the \gls{ETM} are at least as good as the simplex-constrained
$\Delta$-\gls{NVDM}.

These figures show that, of the interpretable models, the \gls{ETM}
provides the best predictive performance while keeping interpretable
topics.  It is robust to large vocabularies.

\subsection{Stop words}
\label{subsec:experiments_stopwords}

We now study a version of the \emph{New York Times} corpus that
includes all stop words. We remove infrequent words to form a vocabulary
of size $10{,}283$.  Our goal is to show that the labeled \gls{ETM}
provides interpretable topics even in the presence of stop words,
another regime where topic models typically fail. In particular, given that
stop words appear in many documents, traditional topic models learn topics
that contain stop words, regardless of the actual semantics of the topic.
This leads to poor topic interpretability.

We fit \gls{LDA}, the $\Delta$-\gls{NVDM}, and the labeled \gls{ETM}
with $K=300$ topics. (We do not report the \gls{NVDM} because it does
not provide interpretable topics.)  Table\nobreakspace \ref {tab:tc_td_stopwords} shows
topic quality (the product of topic coherence and topic
diversity). Overall, the labeled \gls{ETM} gives the best performance
in terms of topic quality.

\begin{figure}[t]
 \centering
 \includegraphics[width=0.9\linewidth]{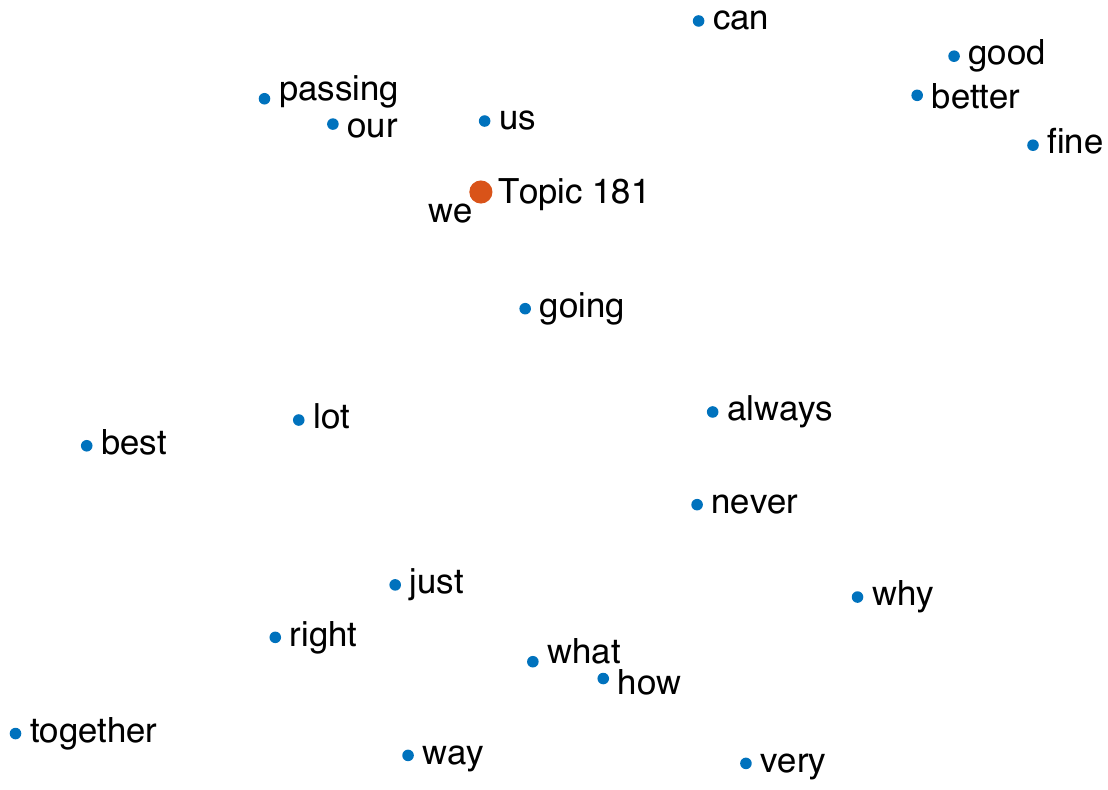}
 \captionof{figure}{A topic containing stop words found by the \acrshort{ETM} on \emph{The New York Times}. The \gls{ETM} is robust even in the presence of stop words.}
 \label{fig:topic_embedding_stops}
\end{figure}

While the \gls{ETM} has a few ``stop topics'' that are specific for
stop words (see, e.g., Figure\nobreakspace \ref {fig:topic_embedding_stops}),
$\Delta$-\gls{NVDM} and \gls{LDA} have stop words in almost every
topic.  (The topics are not displayed here for space constraints.) The
reason is that stop words co-occur in the same documents as every
other word; therefore traditional topic models have difficulties
telling apart content words and stop words. The labeled \gls{ETM}
recognizes the location of stop words in the embedding space;
its sets them off on their own topic.

 \section{Conclusion}
\label{sec:conclusion}

We developed the \gls{ETM}, a generative model of documents that
marries \gls{LDA} with word embeddings. The \gls{ETM} assumes that
topics and words live in the same embedding space, and that words are
generated from a categorical distribution whose natural parameter is
the inner product of the word embeddings and the embedding of the
assigned topic.

The \gls{ETM} learns interpretable word embeddings and topics, even in
corpora with large vocabularies. We studied the performance of the
\gls{ETM} against several document models. The \gls{ETM} learns both
coherent patterns of language and an accurate distribution of words.

\begin{table}[t]
  \centering
  \caption{Topic quality on the \emph{New York Times} data in the
    presence of stop words. Topic quality is the
    product of topic coherence and topic diversity (higher is
    better). The labeled \gls{ETM} is robust to stop words; it achieves
    similar topic coherence than when there are no stop words.
    \label{tab:tc_td_stopwords}}
  {\small
    \begin{tabular}{cccc} \toprule
      & Coherence  &  Diversity  & Quality\\ \midrule
      \acrshort{LDA}           &  $0.13$ & $0.14$ & $0.0173$  \\
            $\Delta$-\acrshort{NVDM} &   $0.17$        &    $0.11$   & $0.0187$    \\
      Labeled \acrshort{ETM}    & $\mathbf{0.18}$ & $\mathbf{0.22}$ & $\mathbf{0.0405}$\\ \bottomrule
    \end{tabular}}
\end{table}

\section*{Acknowledgments}
This work is funded by ONR N00014-17-1-2131, NIH 1U01MH115727-01, DARPA SD2 FA8750-18-C-0130, ONR N00014-15-1-2209, NSF CCF-1740833, the Alfred P. Sloan Foundation, 2Sigma, Amazon, and NVIDIA. FJRR is funded by the European Union's Horizon 2020 research and innovation programme under the Marie Sk\l{}odowska-Curie grant agreement No.\ 706760. 
ABD is supported by a Google PhD Fellowship.

\bibliographystyle{acl_natbib}
\bibliography{main}

\end{document}